\documentclass[preprintnumbers,article,amsmath,amssymb,floatfix,10pt,prd,onecolumn,
superscriptaddress,nofootinbib]{revtex4}
\usepackage[colorlinks=true, pdfstartview=FitV, linkcolor=blue, citecolor=red, urlcolor=magenta]{hyperref}
\usepackage{bbm}
\usepackage{amsfonts}
\usepackage{mathrsfs}
\usepackage{latexsym}
\usepackage{epsfig}
\usepackage{epstopdf}
\usepackage{epstopdf}
\usepackage{graphicx}
\usepackage{amssymb}
\usepackage{amsmath}
\usepackage{dcolumn}
\usepackage{bm}
\usepackage{color}
\usepackage{comment}
\usepackage{xcolor}
\begin{document}
\title{Gravitational Analysis of Einstein-Non-Linear-Maxwell-Yukawa Black Hole under the Effect of Newman-Janis Algorithm}
\author{Rimsha Babar}
\email{rimsha.babar10@gmail.com} \affiliation{Division of Science
and Technology, University of Education, Township, Lahore-54590,
Pakistan}

\author{Muhammad Asgher}
\email{m.asgher145@gmail.com} \affiliation{Department of
Mathematics, The Islamia University of Bahawalpur, Bahawalpur-63100,
Pakistan}

\author{Riasat Ali}
\email{Correspondence: riasatyasin@gmail.com}
\affiliation{Department of Mathematics, GC, University Faisalabad
Layyah Campus, Layyah-31200, Pakistan}

\begin{abstract}
In this paper, we analyze the rotating
Einstein-non-linear-Maxwell-Yukawa black hole solution by
Janis-Newman algorithmic rule and complex calculations. We
investigate the basic properties (i.e., Hawking radiation) for the
corresponding black hole solution. From the horizon structure of the
black hole, we discuss the graphical behavior of Hawking temperature
$T_H$ and analyze the effects of spin parameter (appears due to
Newman-Janis approach) on the $T_H$ of black hole. Furthermore, we
investigate the corrected temperature for rotating
Einstein-non-linear-Maxwell-Yukawa black hole by using the vector
particles tunneling strategy which is based on Hamilton-Jacobi
method. We additionally study the graphical explanation of corrected
$T_H$ through outer horizon to investigate the physical and stable
conditions of black hole. Finally, we compute the corrected entropy
and check that the effect of charged, rotation and gravity on
entropy.\\\\{\bf Keywords:}Yukawa Black Hole; Newman-Janis algorithm;
Semi-classical Phenomenon; Hawking temperature; corrected entropy.\\
\end{abstract}

\maketitle

\date{\today}

%%%%%%%%%%%%%%%%%%%%%%%%%%%%%%%%%%%%%%%%%%%%%%%%%%%%%%%%%%%%%%%%%%%%%%%%
%%%%%%%%%%%%%%%        Introduction        %%%%%%%%%%%%%%%%%%%%%%%%%%%%%
%%%%%%%%%%%%%%%%%%%%%%%%%%%%%%%%%%%%%%%%%%%%%%%%%%%%%%%%%%%%%%%%%%%%%%%%

\section{Introduction}
One of the main feature of Einstein's theory of general relativity
(GR) is the well-known conservation of the covariance of the energy
momentum tensor (EMT) that according to the Noether symmetry
postulate, prompts to a globally defined transformation of physical
geometry. These moderated amounts show up in the form of integrals
of the parts of the EMT on the corresponding spatial surface. These
space-like surfaces concede at any rate some particular Killing
vectors of the basic space-time as their typical. Therefore, the
rest mass/total energy of the system is conserved with regards to
GR. After this, some new generalized theories related about GR have
been suggested that weaken the state of conservation of EMT. One of
the feasible change of the overall theories of GR was presented by
P. Rastall ($1972$) \cite{[1], [2]}. The standard law of
conversation can be determined from null divergence
($T^{uv}_{;u}=0$). At that point, the non-minimal connection of the
matter with the geometry is considered in which the divergence
($T_{uv}$) is proportional to the Ricci scalar gradient
($T^{uv}_{;u}\propto R^{,v}$), so that, the standard conservation
law is recovered within the flat space-time. The Rastall theory
\cite{[18]} may be an understandable theory within the idea that the
fluctuation of any amplified gravity theory from the basic GR should
be frail to overcome the sun oriented system evaluation. A few
examinations on the different parts of this theory with regards to
current speed up expansion period of the universe also other
cosmological issues can be found \cite{[12]}. The substance of the
Rastall theory is related to the higher curvature situations and
thus the black holes (BHs) physics give a suitable area in arrange
to examine this theory in more subtleties. Other fascinating issues
are that Rastall gravity appears to don't endure from the entropy as
well as age issues of usual cosmology \cite{[23]} and it is also
predictable with the gravitational lensing wonders \cite{[24]}. Many
other investigations on Rastall theory have been studied and
references \cite{[27]}-\cite{[30]}.

The tunneling phenomenon is observed for boson charged particles
with electro-positive energy crosses the horizon of BH and these
particles appear as Hawking radiation \cite{R4,RA}. The Hawking
radiation and entropy of BH by applying the tunneling mechanism to
the non-spherical Kerr and Kerr-Newman metric has been investigated
\cite{R7}. The author calculated the entropy effects on the
stability of BH.

The tunneling probability inside and outside the horizon as well as
$T_H$ by taking into account the Newman-Penr\"{o}se formalism and
Hamilton Jacobi-method has been studied \cite {R8}. From their
analysis, they concluded that the $T_H$ depends upon mass of a BH,
magnetic charge and electric charge. The $T_H$ by incorporating
fermions tunneling from squashed BH in the G\"{o}del universe and
charged Kaluza-Kliein spacetime has been investigated \cite{R9}. The
author showed that the fermion and scalar particles with spin tunnel
through the horizon give identical expressions for $T_H$. By taking
into account the global mono-pole charge, Sharif and his fellow
\cite{R12} have studied the Hawking radiation from
Reissner-Nordstr\"{o}m de Sitter BH. They extended their analysis by
studying back reaction effects of fermions tunneling through
horizon. The same authors also studied the thermodynamics of
different BH solutions to the fermions. They also evaluated surface
gravity, $T_H$ and first law of thermodynamics \cite{R13}. The
tunneling radiation phenomenon was also discussed by Silva and Brito
\cite{R14}. The advantage of discussing tunneling phenomenon is
that, it is helpful to remove singularity as well to discuss
thermodynamical properties of BHs. They also calculated the emission
spectrum of self-dual BH. Quantum tunneling for Hawking radiation
via static and dynamic BH was carried by Chakraborty and Saha
\cite{R15}. They calculated quantum corrections upon the first order
as the equations of motions for higher orders were complicated. They
also discussed law of BH mechanics considering modified $T_H$.
Darabi et al. \cite{R16} studied the Hawking radiations from
generalized rotating and static BH. They have made use of WKB
approximation for calculating tunneling. Liu \cite{R17} have
analyzed the energy conservation of charged particles as well as
quantum corrections via tunneling method for a modified version of
Reissner BH. He concluded that the entropy is independent from the
dispersion relations over matter fields. Ding et al. \cite{R18}
studied the tunneling process of relativistic and non-relativistic
particles for Killing and universal horizon of BH by using
Hamilton-Jacobi process. They calculated the entropy effect on
stability of BH.

The Hawking radiation and classical tunneling by applying the WKB
approximation and the ray phase space process has been analyzed
\cite{R20}. Moreover, they also determined the outgoing charged
particles tunneling probability at event horizon of BH. Jusufi et
al. \cite{R21} obtained the Hawking radiation of vector particle by
applying the WKB-approximation in Friedman-Robertson-Walker (FRW)
universe in a BH. They calculated the effects of gravity by
radiation phenomenon of charged boson particle of BH and black ring.
The BH radiation with modified dispersion relation in tunneling
paradigm statical frame has been investigated \cite{R22}. The
authors in \cite{R23} have calculated the geodesic equations for
massive/massless particles in a very effective way. They also
studied the radiation process with the help of tunneling strategy
from cosmic BH. The particle dynamics around the Kerr MOG BH with
magnetic field has been discussed. Sharif and Shahzadi \cite{R24}
concluded that the external magnetic field has strong influence on
particle dynamics in MOG depending upon the spin of a BH. Javed and
his colleagues \cite{R25}-\cite{B} have computed the charged boson
particle tunneling with the help of couple of accelerating/rotating
super gravity BH in $5D$. In particular, they have used WKB
approximation to study tunneling and Hawking temperature.
Cvetkovi\'{c} and Simic \cite{R27} investigated the static
spherically symmetric solutions of Lovelock gravity by using torsion
process. They found well-known solution, which is known as
Boulware-Deser BH. \"{O}vg\"{u}n et al. \cite{R28} investigated the
$T_H$ of BH by calculating the tunneling rate of massive charge
vector particles with the electromagnetic field.

The Hawking radiation process via tunneling method for $5$D
Myers-Perry BH has been investigated \cite{R29}. Generalized
uncertainty principle (GUP) effects on temperature via geometry of
BH was examined by Gecim and his fellow \cite{R30}. For this aim,
the authors concluded that the $T_H$ increases with the increase of
angular momentum. Javed and Babar \cite{R31} discussed charged
fermion particles tunneling via Kerr-Newman-Ads BH. In order to
study the required task, they used Hamilton-Jacobi anstaz to
calculate the $T_H$ for spin-$1/2$ particles. Ali et al.
\cite{R32}-\cite{T6} investigated the field equation for massive
boson via WKB method and also discussed the gravity effects on
radiation to check the stability and instability of BH.

The basic intention of this article is to derive the
Einstein-non-linear Maxwell-Yukawa (ENLMY) BH in the background of
Newman-Janis algorithm and to extend the ENLMY BH into rotating
ENLMY BH. Furthermore, to discuss the stability conditions of RENLMY
BH via graphical interpretation of its $T_H$. The paper is formatted
in the following manner: In section  \textbf{II}, we discuss a
RENLMY BH solution in the Newman-Janis algorithm and also analyze
the $T_H$ for the corresponding BH. The section \textbf{III}
comprises the graphical analysis of $T_H$ with horizon and check the
stable condition of RENLMY BH. The section \textbf{IV} study the
$T'_{H}$ (corrected temperature) for RENLMY BH. The section
\textbf{V} analyzes quantum gravity effects on RENLMY BH with
graphical evaluation. Section \textbf{VI} study the corrected
entropy for RENLMY BH and its graphical analysis. At last, Sec.
\textbf{VII} contains the summary and conclusions.

\section{Rotating Einstein-Non-Linear-Maxwell-Yukawa black hole}
By utilizing the Newman-Janis method, the rotation parameter $a$ can
be calculated in a spherically symmetric explanation that gives the
modification of Newman-Janis method. Here, we derive a metric for
ENLMY with a modification of rotation parameter by utilizing
Newman-Janis method. Furthermore, we compute the $T_H$ for the given
solution of BH . The ENLMY BH with a spherically symmetric static
metric can be written as \cite{R1, RR}

\begin{equation}
ds^{2}=-F(r)dt^2+\frac{1}{F(r)}dr^2+r^2 d\theta^2+r^2 \sin^2 \theta
d\phi^2,\label{a11}
\end{equation}
where $F(r)\simeq
1+\frac{2m}{r}-\frac{QC_{0}}{r^2}+\frac{4QC_{0}\beta}{3r}-Q
C_{0}\beta^{2}+O(\beta^3)$. Here, $m$ represents the mass of BH,
$C_{0}$ is an integration constant(dimensionless parameter), Q
depicts the charge of BH that is located at the origin and $\beta$
is a positive constant and it can be chosen as $\beta = 1$. The
vector potential of the Yukawa black hole can be defined as
\begin{equation}
A=\frac{Q}{r^{2}e^{r}}dt
\end{equation}

We consider the Eddington-Finkelstein (EF) coordinates
transformations $(t, r, \theta, \phi)$ to the Boyer-Lindquist (BL)
coordinates $(u, r, \theta, \phi)$, we get
\begin{eqnarray}
du=dt-\frac{dr}{F(r)}.\label{A}
\end{eqnarray}
and by applying this coordinate transformation to metric Eq.
(\ref{a11}), we have
\begin{equation}
ds^{2}=-F(r)du^2-2dudr+r^2 d\theta^2+r^2 \sin^2\theta
d\phi^2.\label{Q1}
\end{equation}
The components of metric in the background of null framework can be
re-written as
\begin{eqnarray}
g^{\mu\nu}=-l^\nu n^\mu-l^\mu n^\nu+m^\mu \bar{m}^{\nu}+m^\nu
\bar{m}^{\mu}.
\end{eqnarray}
Here the evaluating elements are
\begin{eqnarray}
l^{\mu}&=&\delta_{r}^{\mu},~~~n^{\nu}=\delta_{u}^{\mu}-\frac{1}{2} F \delta_{r}^{\mu},\nonumber\\
m^{\mu}&=&\frac{1}{\sqrt{2}r} \delta_{\theta}^{\mu}+\frac{i}{\sqrt{2}r \sin\theta}\delta_{\phi}^{\mu},\nonumber\\
\bar{m}^{\mu}&=&\frac{1}{\sqrt{2}r}
\delta_{\theta}^{\mu}-\frac{i}{\sqrt{2}r\sin\theta}\delta_{\phi}^{\mu}.\nonumber
\end{eqnarray}
The following relation of the null tetrad are satisfied in $(u, r)$
plane
$l_{\mu}l^{\mu}=n_{\mu}n^{\mu}=m_{\mu}m^{\mu}=l_{\mu}m^{\mu}=m_{\mu}m^{\mu}=0$
and $l_{\mu}n^{\nu}=-m_{\mu}\bar{m}^{\mu}=1$, we can choose the
coordinate transformation as $r\rightarrow r+ia\cos\theta$,
$u\rightarrow u-ia\cos\theta$, then we perform the transformation
$F(r)\rightarrow \tilde{F}(r, a, \theta)$ and
$\Sigma^2=a^2\cos^2\theta+r^2$. The vectors in given space become
\begin{eqnarray}
l^{\mu}&=&\delta_{r}^{\mu},~~~n^{\nu}=\delta_{u}^{\mu}-\frac{1}{2}
\tilde{F} \delta_{r}^{\mu},\\
m^{\mu}&=&\frac{1}{\sqrt{2}r}\left(\delta_{\theta}^{\mu}+(\delta_{u}^{\mu}-\delta_{r}^{\mu})ia \sin\theta+\frac{i}{\sin\theta}\delta_{\phi}^{\mu}\right),\\
\bar{m}^{\mu}&=&\frac{1}{\sqrt{2}r}\left(\delta_{\theta}^{\mu}-(\delta_{u}^{\mu}
-\delta_{r}^{\mu})ia
\sin\theta-\frac{i}{\sin\theta}\delta_{\phi}^{\mu}\right).
\end{eqnarray}
From the definition of the null tetrad the metric tensor $g^{\mu r}$
in the EF coordinate can be given as
\begin{eqnarray}
g^{uu}&=&\frac{a^2\sin\theta^2}{\sum^2},~~~g^{ur}=-1-\frac{a^2\sin\theta^2}{\sum^2},
~~~g^{rr}=\tilde{F}+\frac{a^2\sin\theta^2}{\sum^2},~~~
g^{\theta\theta}=\frac{1}{\sum^2},\nonumber\\
g^{\phi\phi}&=&\frac{1}{\sum^2\sin\theta^2},~~~g^{ur}=\frac{\sin\theta}{\sum^2},~~~
g^{r\phi}=-\frac{a}{\sum^2},\nonumber
\end{eqnarray}
where
\begin{equation}
\tilde{F}(r, \theta)=1-\frac{2mr}{\Sigma^2}-\frac{QC_{0}}{\Sigma^2}
+\frac{4QC_{0}\beta r}{3\Sigma^2}-\frac{QC_{0}\beta^2
r^2}{\Sigma^2}+O(\beta^3).
\end{equation}
We do the coordinate transformation from EF to BL coordinates as
\begin{equation}
du=dt+Y(r)dr,~~~d\phi=d\phi+Z(r)dr,
\end{equation}
where
\begin{equation}
Y(r)=\frac{r^{2}+a^{2}}{r^{2}F
+a^{2}},~~~~~Z(r)=-\frac{a}{r^{2}F+a^{2}}.
\end{equation}
Finally, we get the EF coordinate transformation in the following
from
\begin{eqnarray}
ds^{2}&=&-\frac{\left[r^2-2mr-QC_{0}+\frac{4QC_{0}\beta r}{3}-Q
r^2 C_{0}\beta^{2}+ O(\beta^3)\right]}{\Sigma^2}dt^2+\frac{\Sigma^2}{\Delta_{r}}dr^2\nonumber\\
&-&2a\sin\theta^2\left[1-\frac{r^2-2mr-QC_{0}+\frac{4QC_{0}\beta
r}{3}-Q
r^2 C_{0}\beta^{2}+ O(\beta^3)}{\Sigma^2}\right]dt d\phi+\Sigma^2d\theta^2\nonumber\\
&+&\frac{a^2\sin^2\theta\left[\Sigma^4-a^2\sin^2\theta
\left(r^2-2mr-QC_{0}+\frac{4QC_{0}\beta r}{3}-Q r^2 C_{0}\beta^{2}+
O(\beta^3)-2\right)\right]}{\Sigma^2}d\phi^2.\label{Met}
\end{eqnarray}
where
\begin{eqnarray}
\Delta_{r} =r^2-2mr-QC_{0}+\frac{4QC_{0}\beta r}{3}-Q r^2
C_{0}\beta^{2}+ O(\beta^3).\nonumber
\end{eqnarray}
Thus the Eq. (\ref{Met}) gives the metric for rotating ENLMY BH with
the spin parameter $a$. In order to study the thermodynamical
properties of ENLMY BH in rotating case, we compute the $T_H$ by the
given formula \cite{AA1}
\begin{equation}
T_{H} = \frac{\tilde{F'}(r_+)}{4\pi}.
\end{equation}
The temperature $T_H$ for rotating (RENLMY) BH can be derived as
\begin{eqnarray}
T_{H}&=&\frac{3mr_{+}^{2}-3C_{0}Qr_{+}+2C_{0}Qr_{+}^{2}\beta+a^2(3C_0Q\beta
r_{+} -2C_{0}Q\beta-3m)}{6\pi(r_{+}^{2}+a^2)^2}.\label{th1}
\end{eqnarray}

From our calculation, we have computed that the $T_H$ at which
charged particles radiation both in and out through the horizons $r$
is independent of the types of the particles, and $T_H$ depends upon
mass $m$, BH charge $Q$, spin parameter $a$ and arbitrary constants
$\beta, C_{0}$. It is concerning to note that for $a=0$, we recover
$T_H$ with out Newman-Janis algorithm.

\section{Graphical Analysis of $T_{H}$ for RENLMY BH}
The section give the graphical conduct of $T_H$ with horizon $r_+$
for RENLMY BH. We investigate the effects of spin parameter $a$ and
charge $Q$ of RENLMY BH on $T_{H}$. We see the behavior of $T_{H}$
by setting the fixed value of mass $m=1$ and arbitrary constants
$\beta=0.9=C_{0}$. Moreover, we analyze different parameters on the
stability of RENLMY BH.

\begin{center}
\includegraphics[width=8.4cm]{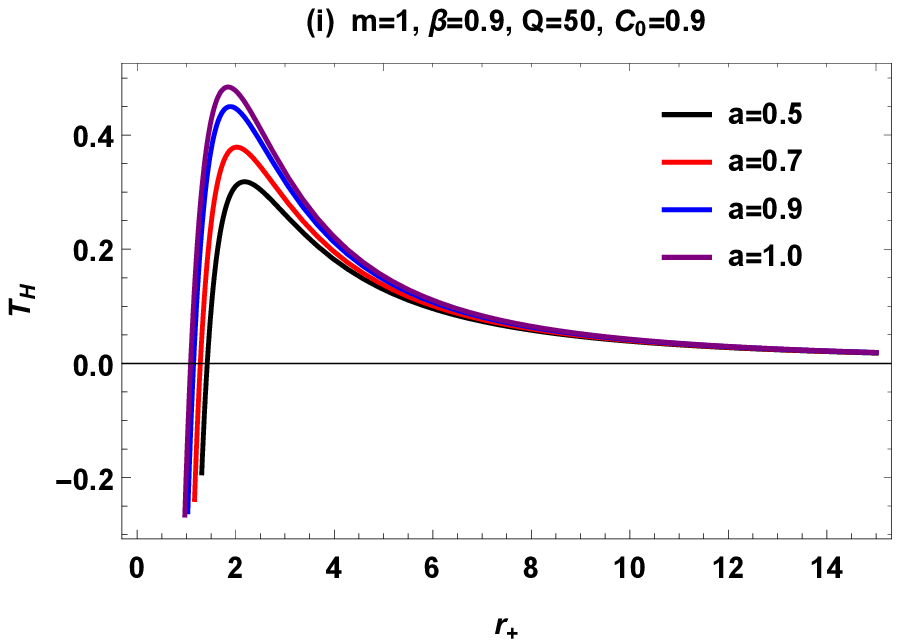}\includegraphics[width=8.4cm]{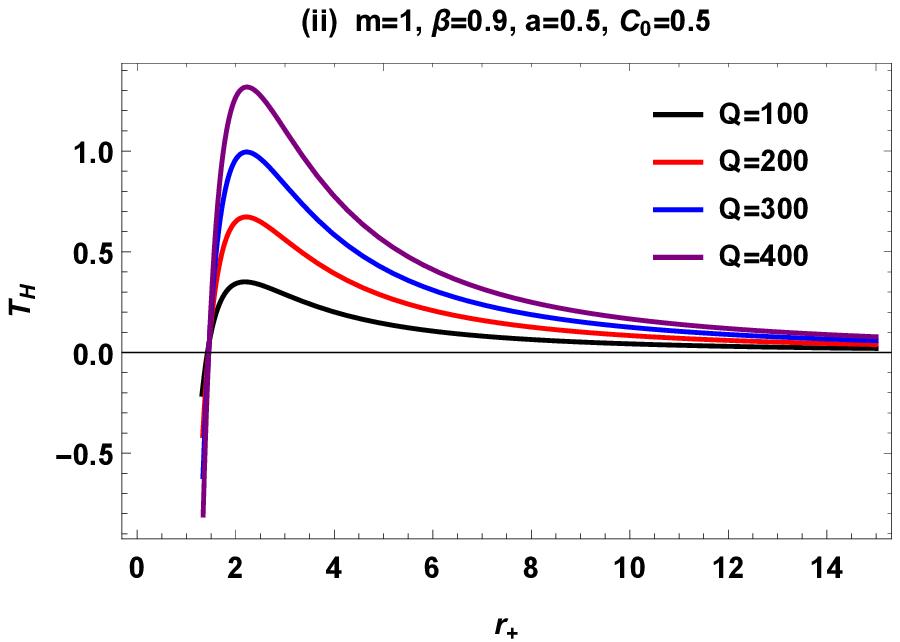}\\
{Figure 1: $T_{H}$ via horizon $r_{+}$}.
\end{center}
\textbf{Figure 1}: (i) states the interpretation of $T_{H}$ for
constant value of charge $Q=50$ and different values of spin
parameter $a$. It is observable that at the initial stage the BH is
unstable (due to negative $T_H$) but as the time passes the BH gets
its stable form. After attaining a maximum height the $T_{H}$
eventually drops down to get an asymptotically flat condition till
$r_{+}\rightarrow\infty$. This condition guarantee the stable state
of BH. We can also observe that the $T_H$ increases with the
increasing value of spin parameter. (ii) shows the interpretation of
$T_{H}$ for different values of RENLMY BH charge $Q$ and constant
spin parameter $a=0.5$. One can see that with the passage of time
the BH becomes stable by getting an asymptotically flat state after
dropping down from height at maximum $T_H$. In this case the $T_H$
increases as we increase the value of charge.

\section{Corrected $T_H$ for RENLMY BH}
In order to check the $T'_{H}$ for RENLMY BH, we follow the vector
tunneling strategy by considering the Hamilton–Jacobi method. In
this section, we explore vector particles tunneling for further
generalized rotating Yukawa BH with rotation parameter and also
analyze the $T'_{H}$ at which the vector particles radiate through
horizon. For this purpose, we use the modified wave equation to
study the tunneling behavior of vector particles from RENLMY BH.

In order to meet our goal, the metric Eq. (\ref{Met}) can be written
as
\begin{eqnarray}
ds^{2}&=&-\tilde{V}dt^{2}+\tilde{W}dr^{2}+\tilde{X}d\theta^{2}
+\tilde{Y} d\phi^{2}+2\tilde{Z}dt d\phi,\nonumber\label{aa}
\end{eqnarray}
where $ \tilde{V}=\tilde{F}(r),~\tilde{W}=-\delta,
~\tilde{X}=\Sigma^2,~~\tilde{Y}
=\sin\theta^2\left[\Sigma^2-a^2\sin\theta^2(\tilde{F}-2)\right]$,
$~\tilde{Z}=2a\left(\tilde{F}(r)-1\right)\sin^2\theta.$

The generalized wave equation for vector particles motion can be
expressed as \cite{R33}
\begin{equation}
\partial_{\mu}(\sqrt{-g}\phi^{\nu\mu})+\sqrt{-g}\frac{m^2}{\hbar^2}\phi^{\nu}
+\sqrt{-g}\frac{i}{\hbar}A_{\mu}\phi^{\nu\mu}
+\sqrt{-g}\frac{i}{\hbar}eF^{\nu\mu}\phi_{\mu}+\alpha\partial_{0}\partial_{0}\partial_{0}(\sqrt{-g}g^{00}\phi^{0\nu})\hbar^{2}
-\alpha
\partial_{i}\partial_{i}\partial_{i}(\sqrt{-g}g^{ii}\phi^{i\nu})\hbar^{2}=0,
\end{equation}
here $g$ is determinant of coefficient matrix, $\phi^{\nu\mu}$ is
anti-symmetric tensor and $m$ is particle mass.

The $\phi_{\nu\mu}r$ tensor can be expressed as
\begin{equation}
\phi_{\nu\mu}=-
(1-\alpha{\hbar^2\partial_{\mu}^2})\partial_{\mu}\phi_{\nu}+(1-\alpha{\hbar^2\partial_{\nu}^2})\partial_{\nu}\phi_{\mu}-(1-\alpha{\hbar^2}\partial_{\nu}^2)\frac{i}{\hbar}eA_{\mu}\phi_{\nu}+(1-\alpha{\hbar^2\partial_{\nu}^2})
\frac{i}{\hbar}eA_{\nu}\phi_{\mu},~~~~~ F_{\nu\mu}=\nabla_{\nu}
A_{\mu}-\nabla_{\mu} A_{\nu},\nonumber
\end{equation}
where $\alpha,~A_{\mu},~e~$ and $\nabla_{\mu}$ are the quantum
gravity parameter (correction parameter), Yukawa BH vector
potential, the particle charge and derivatives of co-variant,
respectively.

According to the above metric the non-zero components can be
calculated as
\begin{eqnarray}
&&\phi^{0}=\frac{\tilde{-Y}\phi_{0}+\tilde{Z}\phi_{3}}{\tilde{V}\tilde{Y}
+\tilde{Z}^2},~~~\phi^{1}=\frac{1}{\tilde{W}}\phi_{1},
~~~\phi^{2}=\frac{1}{\tilde{X}}\phi_{2},~~~
\phi^{3}=\frac{\tilde{Z}\phi_{0}+\tilde{V}\phi_{3}}{\tilde{V}\tilde{Y}
+\tilde{Z}^2},~~~\phi^{01}=\frac{\tilde{-D}\phi_{01}+\tilde{Z}\phi_{13}}
{\tilde{W}(\tilde{V}\tilde{Y}+\tilde{Z}^2)},~~~
\phi^{02}=\frac{\tilde{-Y}\phi_{02}}{\tilde{X}(\tilde{V}\tilde{Y}+\tilde{Z}^2)},\nonumber\\
&&\phi^{03}=\frac{(\tilde{-V}\tilde{Y}+\tilde{V}^2)\phi_{03}}{(\tilde{V}
\tilde{Y}+\tilde{Z}^2)^2},~~~\phi^{12}=\frac{1}{\tilde{W}\tilde{X}}\phi_{12},
~\phi^{13}=\frac{1}{\tilde{W}\tilde{V}\tilde{Y}+\tilde{Z}^2}\phi_{13},
~~\phi^{23}=\frac{\tilde{Z}\phi_{02}+\tilde{V}\phi_{23}}{\tilde{X}(\tilde{V}\tilde{Y}+\tilde{Z}^2)},\nonumber
\end{eqnarray}
The WKB approximation is expressed in the form
\begin{equation}
\phi_{\nu}=c_{\nu}\exp\left[\frac{i}{\hbar}I_{0}(t,r,\theta,\phi)+
\Sigma \hbar^{n}I_{n}(t,r, \theta,\phi)\right].
\end{equation}
The set of field equation are given below
\begin{eqnarray}
&+&\frac{\tilde{Y}}{\tilde{W}(\tilde{V}\tilde{Y}+\tilde{Z}^2)}\Big[c_{1}
(\partial_{0}I_{0})(\partial_{1}I_{0})+\alpha
c_{1}(\partial_{0}I_{0})^{3}
(\partial_{1}I_{0})-c_{0}(\partial_{1}I_{0})^{2}-\alpha
c_{0}(\partial_{1}I_{0})^4
+ c_{1}eA_{0}(\partial_{1}I_{0})\nonumber\\
&+&c_{1}\alpha eA_{0}(\partial_{0}I_{0})^{2}
(\partial_{1}I_{0})\Big]-\frac{\tilde{Z}}{\tilde{W}(\tilde{V}\tilde{Y}+\tilde{Z}^2)}\Big[c_{3}
(\partial_{1}I_{0})^2+\alpha
c_{3}(\partial_{1}I_{0})^4-c_{1}(\partial_{1}I_{0})
(\partial_{3}I_{0})-\alpha c_{1}(\partial_{1}I_{0})(\partial_{3}I_{0})^2\Big]\nonumber\\
&+&
\frac{\tilde{Y}}{\tilde{X}(\tilde{V}\tilde{Y}+\tilde{Z}^2)}\Big[c_{2}
(\partial_{0}I_{0})(\partial_{2}I_{0})\alpha
c_{2}(\partial_{0}I_{0})^3(\partial_{2}I_{0})-c_{0}(\partial_{2}I_{0})^2
-\alpha
c_{0}(\partial_{2}I_{0})^4+c_{2}eA_{0}(\partial_{2}I_{0})+c_{2}eA_{0}\alpha
(\partial_{0}I_{0})^{2}(\partial_{1}I_{0})\Big]\nonumber\\
&+&\frac{\tilde{V}\tilde{Y}}{(\tilde{V}
\tilde{Y}+\tilde{Z}^2)^2}\Big[c_{3}(\partial_{0}I_{0})(\partial_{3}I_{0})\alpha
c_{3}
(\partial_{0}I_{0})^{3}(\partial_{3}I_{0})-c_{0}(\partial_{3}I_{0})^{2}
-\alpha
c_{0}(\partial_{3}I_{0})^4+c_{3}eA_{0}(\partial_{3}I_{0})+c_{3}eA_{0}
(\partial_{0}I_{0})^{2}(\partial_{3}I_{0})\Big]\nonumber\\
&-&m^2\frac{\tilde{Y} c_{0}-\tilde{Z}
c_{3}}{(\tilde{V}\tilde{Y}+\tilde{Z}^2)}=0,
\end{eqnarray}
\begin{eqnarray}
&-&\frac{\tilde{Y}}{\tilde{W}(\tilde{V}\tilde{Y}+\tilde{Z}^2)}\Big[c_{1}(\partial_{0}I_{0})^2
+\alpha
c_{1}(\partial_{0}I_{0})^4-c_{0}(\partial_{0}I_{0})(\partial_{1}I_{0})
-\alpha
c_{0}(\partial_{0}I_{0})(\partial_{1}I_{0})^{3}+c_{1}eA_{0}(\partial_{0}
I_{0})+\alpha c_{1}eA_{0}(\partial_{0}I_{0})^3\Big]\nonumber\\
&+&\frac{\tilde{Z}}{\tilde{W}(\tilde{V}\tilde{Y}+\tilde{Z}^2)}\Big[c_{3}
(\partial_{0}I_{0})(\partial_{1}I_{0})+\alpha
c_{3}(\partial_{0}I_{0})
(\partial_{1}I_{0})^3-c_{1}(\partial_{0}I_{0})(\partial_{3}I_{0})-\alpha
c_{1}
(\partial_{0}I_{0})(\partial_{3}I_{0})^{3}\Big]+\frac{1}{\tilde{W}\tilde{X}}
\Big[c_{2}(\partial_{1}I_{0})(\partial_{2}I_{0})\nonumber\\
&+&\alpha
c_{2}(\partial_{1}I_{0})(\partial_{2}I_{0})^3-c_{1}(\partial_{2}I_{0})^{2}
-\alpha
c_{1}(\partial_{2}I_{0})^{4}\Big]+\frac{1}{\tilde{W}(\tilde{V}\tilde{Y}
+\tilde{Z}^2)}\Big[c_{3}(\partial_{1}I_{0})(\partial_{3}I_{0})+\alpha
c_{3}
(\partial_{1}I_{0})(\partial_{3}I_{0})^3-c_{1}(\partial_{3}I_{0})^2\nonumber\\
&-&\alpha c_{1} (\partial_{3}I_{0})^{4}\Big]-\frac{m^2
c_{1}}{\tilde{W}}
+\frac{eA_{0}\tilde{Y}}{\tilde{W}(\tilde{V}\tilde{Y}+\tilde{Z}^2)}\Big[c_{1}
(\partial_{0}I_{0})+\alpha c_{1}(\partial_{0}I_{0})^3
-c_{0}(\partial_{1}I_{0})-\alpha c_{0}(\partial_{1}I_{0})^3+eA_{0}c_{1}\nonumber\\
&+&\alpha c_{1}eA_{0}(\partial_{0}I_{0})^{2})\Big]
+\frac{eA_{0}\tilde{Z}}{\tilde{W}(\tilde{V}\tilde{Y}+\tilde{Z}^2)}
\Big[c_{3}(\partial_{1}I_{0})+\alpha c_{3}(\partial_{1}I_{0})^3
-c_{1}(\partial_{3}I_{0})-\alpha c_{1}(\partial_{1}I_{0})^3\Big]=0,\\
&+&\frac{\tilde{Y}}{\tilde{X}(\tilde{V}\tilde{Y}+\tilde{Z}^2)}\Big[c_{2}
(\partial_{0}I_{0})^2+\alpha c_{2} (\partial_{0}I_{0})^{4}-c_{0}
(\partial_{0}I_{0})(\partial_{2}I_{0})-\alpha
c_{0}(\partial_{0}I_{0})
(\partial_{2}I_{0})^3+c_{2}eA_{0}(\partial_{0}I_{0})+\alpha
c_{2}eA_{0}
(\partial_{0}I_{0})^{3}\Big]\nonumber\\
&+&\frac{1}{\tilde{W}\tilde{X}}\Big[c_{2}(\partial_{1}I_{0})^2+\alpha
c_{2}
(\partial_{1}I_{0})^{4}-c_{1}(\partial_{1}I_{0})(\partial_{2}I_{0})-\alpha
c_{1}
(\partial_{1}I_{0})(\partial_{2}I_{0})^3\Big]-\frac{\tilde{Z}}{\tilde{X}
(\tilde{V}\tilde{Y}+\tilde{Z}^2)}\Big[c_{2}(\partial_{0}I_{0})(\partial_{3}I_{0})\nonumber\\
&+&\alpha
c_{2}(\partial_{0}I_{0})^{3}(\partial_{3}I_{0})-c_{0}(\partial_{0}I_{0})(\partial_{3}I_{0})
-\alpha c_{0}(\partial_{0}I_{0})^3
(\partial_{3}I_{0})+c_{2}eA_{0}(\partial_{3}I_{0})
+\alpha c_{2}eA_{0}(\partial_{3}I_{0})^{3}\Big]\nonumber\\
&+&\frac{\tilde{V}}{\tilde{X}(\tilde{V}\tilde{Y}+\tilde{Z}^2)}\Big[c_{3}(\partial_{2}I_{0})
(\partial_{3}I_{0})+\alpha
c_{3}(\partial_{2}I_{0})^{3}(\partial_{3}I_{0})-c_{2}
(\partial_{3}I_{0})^2-\alpha c_{2}(\partial_{3}I_{0})^4\Big]-\frac{m^2 c_{2}}{\tilde{X}}\nonumber\\
&+&\frac{eA_{0}\tilde{Y}}{\tilde{X}(\tilde{V}\tilde{Y}+\tilde{Z}^2)}\Big[c_{2}(\partial_{0}I_{0})
+\alpha c_{2}(\partial_{0}I_{0})^3-(\partial_{2}I_{0})c_{0}
-\alpha(\partial_{2}I_{0})^3c_{0}+eA_{0}c_{2}+(\partial_{0}I_{0})^2c_{2}\alpha eA_{0}\Big]=0,\\
&+&\frac{(\tilde{V}\tilde{Y})-\tilde{V}^2}{(\tilde{V}\tilde{Y}+\tilde{Z}^2)^2}
\Big[c_{3}(\partial_{0}I_{0})^2+\alpha
c_{3}(\partial_{0}I_{0})^4-c_{0}
(\partial_{0}I_{0})(\partial_{3}I_{0})-\alpha
c_{0}(\partial_{0}I_{0})
(\partial_{3}I_{0})^{3}+{eA_{0}c_3}(\partial_{0}I_{0})
+\alpha c_{3}eA_{0}(\partial_{0}I_{0})^{3}\Big]\nonumber\\
&-&\frac{\tilde{Y}}{\tilde{X}(\tilde{V}\tilde{Y}+\tilde{Z}^2)}
\Big[c_{3}(\partial_{1}I_{0})^2+\alpha
c_{3}(\partial_{1}I_{0})^{4}-c_{1}
(\partial_{1}I_{0})(\partial_{3}I_{0})-\alpha
c_{1}(\partial_{1}I_{0})
(\partial_{3}I_{0})^3\Big]-\frac{\tilde{Z}}{\tilde{X}
(\tilde{V}\tilde{Y}+\tilde{Z}^2)}\Big[c_{2}(\partial_{0}I_{0})(\partial_{2}I_{0})\nonumber\\
&+&\alpha
c_{2}(\partial_{0}I_{0})^3(\partial_{2}I_{0})-c_{0}(\partial_{2}I_{0})^{2}
+\alpha c_{0}(\partial_{2}I_{0})^4+{eA_{0}c_2}(\partial_{2}I_{0})
+\alpha c_{2}eA_{0}(\partial_{0}I_{0})^{2}(\partial_{2}I_{0})\Big]
-\frac{eA_{0}\tilde{V}}{\tilde{X}(\tilde{V}\tilde{Y}+\tilde{Z}^2)}
\Big[c_{3}(\partial_{2}I_{0})^2\nonumber\\
&+&\alpha
c_{3}(\partial_{2}I_{0})^4-c_{2}(\partial_{2}I_{0})(\partial_{3}I_{0})
-\alpha c_{2}(\partial_{0}I_{0})(\partial_{3}I_{0})^{3}\Big]
-\frac{m^2
(\tilde{Z}c_{0}-\tilde{V}c_{3})}{(\tilde{V}\tilde{Y}+\tilde{Z}^2)}
+\frac{eA_{0}(\tilde{V}\tilde{Y})-\tilde{V}^2}{(\tilde{V}\tilde{Y}+\tilde{Z}^2)^2}
\Big[c_{3}(\partial_{0}I_{0})+\alpha c_{3}(\partial_{0}I_{0})^3\nonumber\\
&-&c_{0}(\partial_{3}I_{0})-(\partial_{3}I_{0})^3\alpha
c_{0}+eA_{0}c_{3} +(\partial_{0}I_{0})^2\alpha eA_{0}\Big]=0,
\end{eqnarray}
By applying the variables separation technique, we can take
\begin{equation}
I_{0}=-(E-j\omega)t+W(r)+\nu(\theta)+J\phi,
\end{equation}
here $\omega$ \& $J$ stands for angular momentums of BH and radiated
particles, respectively, whereas $\tilde{E}=E-j\omega$ is the
particle energy.
\begin{equation*}
K(c_{0},c_{1},c_{2},c_{3})^{T}=0,
\end{equation*}
which gives a $4\times4$ matrix"$K$", whose elements are given as
follows
\begin{eqnarray}
K_{00}&=&\frac{\tilde{-D}}{\tilde{W}(\tilde{V}\tilde{Y}+\tilde{Z}^2)}[W_{1}^2+\alpha
W_{1}^4]-\frac{\tilde{Y}}{\tilde{X}(\tilde{V}\tilde{Y}+\tilde{Z}^2)}[J^2+\alpha
J^4]
-\frac{\tilde{V}\tilde{Y}}{(\tilde{V}\tilde{Y}+\tilde{Z}^2)^2}[\nu_{1}^2+
\alpha \nu_{1}^4]-\frac{m^2 \tilde{Y}}{(\tilde{V}\tilde{Y}+\tilde{Z}^2)},\nonumber\\
K_{01}&=&\frac{\tilde{-D}}{\tilde{W}(\tilde{V}\tilde{Y}+\tilde{Z}^2)}[\tilde{E}
+\alpha \tilde{E}^3+eA_{0}+\alpha eA_{0}\tilde{E}^2]W_{1}
+\frac{\tilde{Z}}{\tilde{W}(\tilde{V}\tilde{Y}+\tilde{Z}^2)}+[\nu_{1}+
\alpha \nu_{1}^3],\nonumber\\
K_{02}&=&\frac{\tilde{-D}}{\tilde{X}(\tilde{V}\tilde{Y}+\tilde{Z}^2)}[\tilde{E}
+\alpha \tilde{E}^3-eA_{0}-\alpha eA_{0}\tilde{E}^2]J,\nonumber\\
K_{03}&=&\frac{\tilde{-E}}{\tilde{W}(\tilde{V}\tilde{Y}+\tilde{Z}^2)}[W_{1}^2+\alpha
W_{1}^4]-
\frac{\tilde{V}\tilde{Y}}{\tilde{X}(\tilde{V}\tilde{Y}+\tilde{Z}^2)^2}[\tilde{E}+\alpha
\tilde{E}^3
-eA_{0}-\alpha eA_{0}\tilde{E}^2]\nu_{1}+\frac{m^2\tilde{Z}}{(\tilde{V}\tilde{Y}+\tilde{Z}^2)^2},\nonumber\\
K_{10}&=&\frac{\tilde{-D}}{\tilde{W}(\tilde{V}\tilde{Y}+\tilde{Z}^2)}[\tilde{E}W_{1}+\alpha
\tilde{E}W_{1}^3] -\frac{m^2}{\tilde{W}}
-\frac{eA_{0}\tilde{Y}}{\tilde{W}(\tilde{V}\tilde{Y}+\tilde{Z}^2)}[W_{1}+\alpha
W_{1}^3],\nonumber
\end{eqnarray}
\begin{eqnarray}
K_{11}&=&\frac{\tilde{-D}}{\tilde{W}(\tilde{V}\tilde{Y}+\tilde{Z}^2)}[\tilde{E}^2
+\alpha \tilde{E}^4-eA_{0}\tilde{E}-\alpha eA_{0}\tilde{E}W_{1}^2]
+\frac{\tilde{Z}}{\tilde{W}(\tilde{V}\tilde{Y}+\tilde{Z}^2)}+[\nu_{1}+
\alpha \nu_{1}^3]\tilde{E}-\frac{1}{\tilde{W}\tilde{X}}[J^2+\alpha J^4]\nonumber\\
&-&\frac{1}{\tilde{W}(\tilde{V}\tilde{Y}+\tilde{Z}^2)}[\nu_{1}+\alpha
\nu_{1}^3]
-\frac{m^2}{\tilde{W}}-\frac{eA_{0}\tilde{Y}}{\tilde{W}(\tilde{V}\tilde{Y}+\tilde{Z}^2)}
[\tilde{E}+\alpha \tilde{E}^3-eA_{0}-\alpha eA_{0}\tilde{E}^2]
+\frac{eA_{0}\tilde{Z}}{\tilde{W}(\tilde{V}\tilde{Y}+\tilde{Z}^2)}[\nu_{1}+
\alpha \nu_{1}^3],\nonumber\\
K_{12}&=&\frac{1}{\tilde{W}\tilde{X}}[W_{1}+\alpha
W_{1}^3]J,~~~~~~~~
K_{21}=\frac{1}{\tilde{W}\tilde{X}}[J+\alpha J^3]W_{1},\nonumber\\
K_{13}&=&\frac{\tilde{-E}}{\tilde{W}(\tilde{V}\tilde{Y}+\tilde{Z}^2)}[W_{1}+\alpha
W_{1}^3]\tilde{E}+
\frac{1}{\tilde{W}(\tilde{V}\tilde{Y}+\tilde{Z}^2)^2}[W_{1}+\alpha
W_{1}^3]\nu_{1}
+\frac{\tilde{Z}eA_{0}}{\tilde{W}(\tilde{V}\tilde{Y}+\tilde{Z}^2)}[W_{1}+\alpha W_{1}^3],\nonumber\\
K_{20}&=&\frac{\tilde{Y}}{\tilde{X}(\tilde{V}\tilde{Y}+\tilde{Z}^2)}[\tilde{E}J+\alpha
\tilde{E}J^3]+
\frac{\tilde{Z}}{\tilde{X}(\tilde{V}\tilde{Y}+\tilde{Z}^2)}[\tilde{E}
+\alpha
\tilde{E}^3\nu_{1}^2]-\frac{\tilde{Y}eA_{0}}{\tilde{X}(\tilde{V}\tilde{Y}+\tilde{Z}^2)}
[J+\alpha J^3]\nonumber\\
K_{22}&=&\frac{\tilde{Y}}{\tilde{X}(\tilde{V}\tilde{Y}+\tilde{Z}^2)}[\tilde{E}^2
+\alpha \tilde{E}^4-eA_{0}\tilde{E}-\alpha
eA_{0}\tilde{E}]-\frac{1}{\tilde{W}\tilde{X}}
+\frac{\tilde{Z}}{\tilde{X}(\tilde{V}\tilde{Y}+\tilde{Z}^2)}[\tilde{E}
+\alpha \tilde{E}^3-eA_{0}-\alpha eA_{0}\tilde{E}^2]\nu_{1}\nonumber\\
&-&\frac{\tilde{V}}{\tilde{X}(\tilde{V}\tilde{Y}+\tilde{Z}^2)}[\nu_{1}^2+
\alpha
\nu_{1}^4]-\frac{m^2}{\tilde{X}}-\frac{eA_{0}\tilde{Y}}{\tilde{X}
(\tilde{V}\tilde{Y}+\tilde{Z}^2)}[\tilde{E}+\alpha
\tilde{E}^3-eA_{0}
-\alpha eA_{0}\tilde{E}^2],\nonumber\\
K_{23}&=&\frac{\tilde{V}}{\tilde{X}(\tilde{V}\tilde{Y}+\tilde{Z}^2)}
[J+\alpha J^3]\nu_{1},\nonumber\\
K_{30}&=&\frac{(\tilde{V}\tilde{Y}-\tilde{A^2})}{(\tilde{V}\tilde{Y}+\tilde{Z}^2)^2}
[\nu_{1}+\alpha \nu_{1}^3]\tilde{E}+
\frac{\tilde{Z}}{\tilde{X}(\tilde{V}\tilde{Y}
+\tilde{Z}^2)}[J^2+\alpha
J^4]-\frac{m^2\tilde{Z}}{(\tilde{V}\tilde{Y}+\tilde{Z}^2)}
-\frac{eA_{0}(\tilde{V}\tilde{Y}-\tilde{A^2})}{(\tilde{V}\tilde{Y}
+\tilde{Z}^2)^2}[\nu_{1}+\alpha \nu_{1}^3],\nonumber\\
K_{31}&=&\frac{1}{\tilde{W}(\tilde{V}\tilde{Y}+\tilde{Z}^2)}[\nu_{1}+\alpha \nu_{1}^3]W_{1},\nonumber\\
K_{32}&=&\frac{\tilde{Z}}{\tilde{X}(\tilde{V}\tilde{Y}+\tilde{Z}^2)}[J+\alpha
J^3]\tilde{E}+
\frac{\tilde{V}}{\tilde{X}(\tilde{V}\tilde{Y}+\tilde{Z}^2)}[\nu_{1}+\alpha \nu_{1}^3]J,\nonumber\\
K_{33}&=&\frac{(\tilde{V}\tilde{Y}-\tilde{A^2})}{(\tilde{V}\tilde{Y}+\tilde{Z}^2)}[\tilde{E}^2
+\alpha \tilde{E}^4-eA_{0}\tilde{E}-\alpha eA_{0}\tilde{E}^3]
-\frac{1}{\tilde{W}(\tilde{V}\tilde{Y}+\tilde{Z}^2)}[W_{1}^2+\alpha
W_{1}^4]
-\frac{\tilde{V}}{\tilde{X}(\tilde{V}\tilde{Y}+\tilde{Z}^2)}[J^2+\alpha J^4]\nonumber\\
&-&\frac{m^2 \tilde{V}}{(\tilde{V}\tilde{Y}+\tilde{Z}^2)}
-\frac{eA_{0}(\tilde{V}\tilde{Y}-\tilde{A^2})}{(\tilde{V}\tilde{Y}+\tilde{Z}^2)}[\tilde{E}
-\tilde{E}^2eA_{0}+\tilde{E}^3\alpha],\nonumber
\end{eqnarray}
where $J=\partial_{\phi}I_{0}$,~~
$\nu_{1}=\partial_{\theta}{I_{0}}$,~~ $W_{1}=\partial_{r}{I_{0}}$
and $\tilde{E}=(E-j\omega)$. To get the non-trivial answer, we set
the det$\textbf{(K)}=0$, so we obtain
\begin{eqnarray}\label{a1}
ImW^{\pm}&=&\pm \int\sqrt{\frac{(\tilde{E}-A_{0}e)^{2}
+X_{1}\Big[1+\alpha\frac{X_{2}}{X_{1}}\Big]}{\frac{(\tilde{V}\tilde{Y}+\tilde{Z}^2)}
{\tilde{W}\tilde{Y}}}}dr,\nonumber\\
&=&\pm
i\pi\frac{(\tilde{E}-A_{0}e)}{2\kappa(r_{+})}\Big[1+\alpha\Xi\Big],
\end{eqnarray}
where $\Xi=6\left(m^{2}+\frac{\left(v^{2}_{1}
+J^{2}_{\phi}\csc^{2}\theta\right)}{r_{+}^{2}} \right)>0$ and
\begin{eqnarray}
X_{1}&=&\frac{\tilde{W}\tilde{Z}}{(\tilde{V}\tilde{Y}+\tilde{Z}^2)}[\tilde{E}
-eA_{0}]\nu_{1}+\frac{\tilde{V}\tilde{W}}{(\tilde{V}\tilde{Y}+\tilde{Z}^2)}
[\nu_{1}^2]-\tilde{W}m^2,\nonumber\\
X_{2}&=&\frac{\tilde{W}\tilde{Y}}{(\tilde{V}\tilde{Y}+\tilde{Z}^2)}[\tilde{E}^4
-2eA_{0}\tilde{E}^3+(eA_{0})^2\tilde{E}^2]
+\frac{\tilde{W}\tilde{Z}}{\tilde{X}(\tilde{V}\tilde{Y}+\tilde{Z}^2)}[\tilde{E}^3-eA_{0}\tilde{E}^2]\nu_{1}
-\frac{\tilde{V}\tilde{W}}{(\tilde{V}\tilde{Y}+\tilde{Z}^2)}[\nu_{1}^4]-W_{1}^4.\nonumber
\end{eqnarray}
The particles tunneling from horizon is defined as
\begin{equation}
\Gamma=\frac{\Gamma_{emission}}{\Gamma_{absorption}}=
\exp\Big[{-2\pi}\frac{(\tilde{E}-A_{0}e)}
{\kappa(r_{+})}\Big]\Big[1+\alpha\Xi\Big].
\end{equation}
here
\begin{equation}
\kappa(r_{+})=\frac{3mr_{+}^{2}-3C_{0}Qr_{+}+2C_{0}Qr_{+}^{2}\beta+a^2(3C_0Q\beta
r_{+} -2C_{0}Q\beta-3m)}{3\pi(r_{+}^{2}+a^2)^2}.
\end{equation}
The $T'_{H}$ of RENLMY BH is calculated by utilizing the Boltzmann
formula \cite{R25}
$\Gamma_{B}=\exp\left[(\tilde{E}-eA_{0})/T'_{H}\right]$ in the form
\begin{eqnarray}
T'_{H}=\frac{3mr_{+}^{2}-3C_{0}Qr_{+}+2C_{0}Qr_{+}^{2}\beta+a^2(3C_0Q\beta
r_{+}
-2C_{0}Q\beta-3m)}{6\pi(r_{+}^{2}+a^2)^2}\Big[1-\alpha\Xi\Big].\label{th2}
\end{eqnarray}
The $T'_{H}$ of RENLMY BH depends upon gravity parameter $\alpha$,
BH charge $Q$, particle mass $m$, spin parameter $a$.

It has worth to mention here that for $\alpha=0$, we recover the
temperature for Eq. (\ref{th1}), further when $a=0$, the $T'_{H}$
reduced into ENLMY BH and when $a=0=Q$ the $T'_{H}$ explicitly
converts to the Schwarzschild BH temperature at ($r_{+}=2m$)
\cite{[25]}. The temperature in terms of residual mass can be
defined as
\begin{equation}
T'_{H}=\frac{1}{8\pi m}\left[
1-6\beta\left(m^{2}+\frac{\left(v^{2}_{1}
+J^{2}_{\phi}\csc^{2}\theta\right)}{r_{+}^{2}} \right)\right],
\end{equation}
where $\omega_{e}=\left(m^{2}+\frac{\left(v^{2}_{1}
+J^{2}_{\phi}\csc^{2}\theta\right)}{r_{+}^{2}} \right)$ gives the
component of kinetic energy for emitted particles. The gravity
corrections reduce the increment in temperature throughout the
radiation method. Due to the quantum corrections, the radiation
stops at few particular temperature and a remnant mass left. A
remnant is a last state of BH evaporation at a very small horizon.
The increment in the temperature ceased whenever this condition
satisfies \cite{[25a]}
\begin{equation}
m\simeq(m-dm)(1+\alpha\Xi),
\end{equation}
whereas $\alpha=\frac{\alpha_0}{M_{p}^2}$ and $dm=\omega_{e}$ as
well as $\omega_{e} \simeq M_{p}$ here $M_{p}$ represents the
Planck's mass and $\alpha_0$ stands for dimensionless parameter
incorporating gravity effects with $\alpha_0<10^{5}$
\cite{[25b],[25c]} and
\begin{equation}
M_{Res} \simeq \frac{M_{p}^2}{\alpha_0\omega_{e}}\gtrsim
\frac{M_{p}}{\alpha_0},
~~~~~~~~~~~T_{Res}\lesssim\frac{\alpha_0}{8\pi M_p}.
\end{equation}
It is note worthy that the value of the corrected temperature
$T'_{H}$ is lesser than the original temperature (without
corrections) and the radiation process ceased into BH, when the BH
gets its minimum amount $M_{Res}$. There are a couple of
inspirations for BH remnants, one of them is that leftovers mass
prevent BH from becoming so hot during the last phase of the
evaporation. Chen and his colleagues have discussed in detail about
the BH remnant in terms of information paradox as well as dark
matter \cite{[25d],[26a]}.

\section{Graphical Analysis of $T'_{H}$ for RENLMY BH}
This section describes the effects of different parameters on
$T'_{H}$ for RENLMY BH. We check the stability condition for RENLMY
BH under quantum gravity effects by setting the values fixed for
mass $m=1$, arbitrary parameter $\Xi=1$. In these plots, the
Hawking's strategy (the BH radius size reduces with the emission of
high radiations) is clearly visible. We can observe in both plots
when the $T'_{H}$ is at its high value the horizon is very small.
This condition states the physical form of BH and ensures its
stability.

\begin{center}
\includegraphics[width=8.4cm]{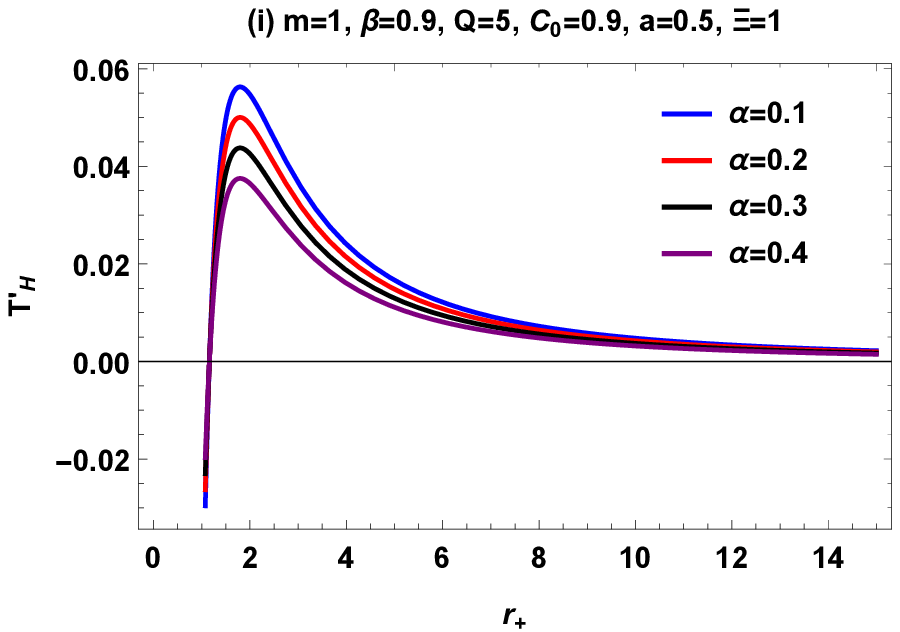}\includegraphics[width=8.4cm]{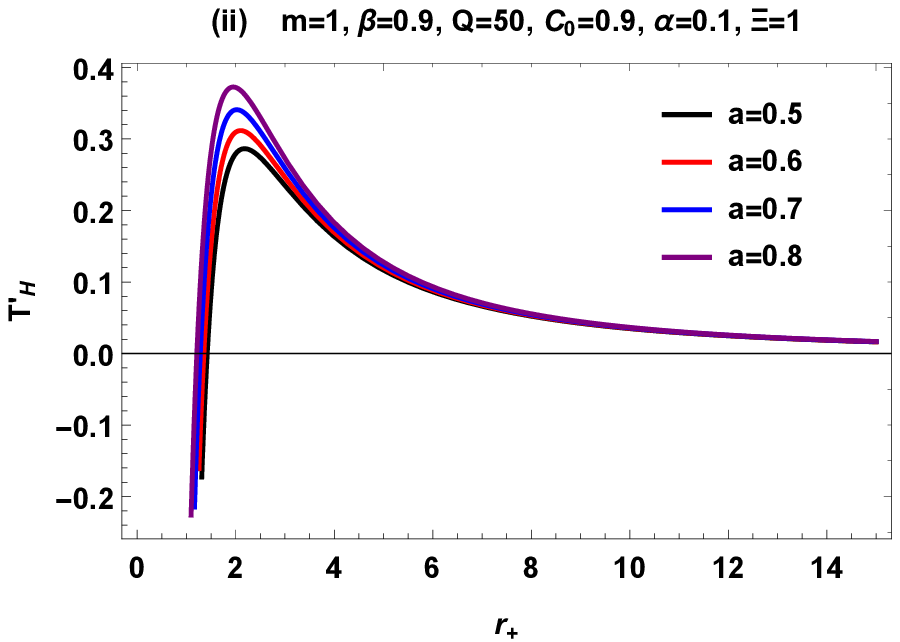}\\
{Figure 2: $T'_{H}$ via horizon $r_{+}$}.
\end{center}

\textbf{Figure 2}: (i) depicts the conduct for $T'_{H}$ with horizon
$r_{+}$ in the region $0\leq r_{+}\leq 15$ for fixed values of
arbitrary constants $\beta=0.9=C_{0}$, spin parameter $a=0.5$, BH
charge $Q=5$ and different values of $\alpha$. One can see that the
BH gets its stable form after passage of some time and attains the
asymptotically flat state upto $r_{+}\rightarrow\infty$. The
$T'_{H}$ increases as we increase the value of gravity parameter
$\alpha$.

(ii) indicates the graphical significance for $T'_{H}$ with horizon
$r_{+}$ various quantities of rotation parameter $a$ and constant
values of arbitrary constants $\beta=0.9=C_{0}$, correction
parameter $\alpha=0.1$, BH charge $Q=50$. One can observe, the
$T'_{H}$ after getting its stable form with positive $T'_{H}$
becomes very high at very small horizon, at this stage the BH
remnant left. Subsequently, attaining this maximum height the
$T'_{H}$ eventually drops down and till $r_{+}\rightarrow\infty$ it
shows an asymptotically flat behavior. This behavior also states the
stable form of BH. We can also check the effects of rotation
parameter on $T'_{H}$, the $T'_{H}$ increases as we increase the
values of rotation parameter.

\section{Corrected Entropy for RENLMY BH}
Here, we evaluate the entropy corrections for RENLMY BH. Banerjee et
al, \cite{c3,c4, c5} have investigated the temperature and entropy
corrections by considering back-reaction effects via null geodesic
phenomenon. We compute the entropy corrections for RENLMY BH in the
background of Bekenstein-Hawking entropy technique for lowest order
corrections. We calculate the logarithmic entropy corrections by
utilizing the corrected temperature $T'_{H}$ and basic entropy
$\mathbb{S}_{o}$ for RENLMY BH. By utilizing the given formula, we
can calculate the corrected entropyas follows
\begin{equation}
\mathbb{S}=\mathbb{S}_{o}-\frac{1}{2}\ln\Big|T_H'^2,
\mathbb{S}_{o}\Big|+...~.\label{vv}
\end{equation}
The basic entropy for RENLMY BH can be derived from the given
expression
\begin{equation}
\mathbb{S}_{o}=\frac{A_{r_{+}}}{4},
\end{equation}
where
\begin{eqnarray}
A_{r_{+}}&=&\int_{0}^{2\pi}\int_{0}^{\pi}\sqrt{g_{\theta\theta}g_{\phi\phi}}d\theta d\phi,\nonumber\\
&=&\frac{2\pi\Big[\Big(r_{+}^2+a^2\Big)^2-a^2\Big(C_{o}Q\Big(1-\frac{4\beta
r_{+}}{3}+r_{+}^2\beta^2\Big)+mr_{+}\Big)\Big]}{\Big(r_{+}^2+a^2\Big)}.\label{kk}
\end{eqnarray}
The basic entropy for RENLMY BH can be computed as
\begin{equation}
\mathbb{S}_{o}=\frac{\pi\Big[\Big(r_{+}^2+mr_{+}+a^2\Big)^2-a^2\Big(C_{o}Q\Big(1-\frac{4\beta
r_{+}}{3}+r_{+}^2\beta^2\Big)\Big)\Big]}{2\Big(r_{+}^2+a^2\Big)}.\label{v1}
\end{equation}
By putting the terms from Eq. (\ref{th2}) \& (\ref{v1}) in the Eq.
(\ref{vv}), we obtain the corrected entropy as
\begin{eqnarray}
\mathbb{S}&=&\frac{\pi\Big[\Big(r_{+}^2+a^2\Big)^2-a^2\Big(C_{o}Q\Big(1-\frac{4r\beta}{3}+r_{+}^2\beta^2\Big)+mr_{+}\Big)\Big]}{2\Big(r_{+}^2+a^2\Big)}\nonumber\\
&-&\frac{1}{2}\ln\left|\frac{\left[\Big(3mr_{+}^{2}-3C_{0}Qr_{+}+2C_{0}Qr_{+}^{2}\beta+a^2(3C_0Q\beta
r_{+}
-2C_{0}Q\beta-3m)\Big)\Big(1-\alpha\Xi\Big)\right]^2\zeta}{72\pi\left(r^2_++a^2\right)^5}\right|+...,\label{b2}
\end{eqnarray}
where
\begin{equation}
\zeta=\Big[\Big(r_{+}^2+a^2\Big)^2-a^2\Big(C_{o}Q\Big(1-\frac{4r\beta}{3}+r_{+}^2\beta^2\Big)+mr_{+}\Big)\Big].\nonumber
\end{equation}
The Eq. (\ref{b2}) shows the entropy corrections for RENLMY BH. It
depends upon charge $Q$, rotation parameter $a$, integration
constant $C_{0}$, arbitrary parameters $\beta,~\Xi$ and correction
parameter $\alpha$.

\begin{center}
\includegraphics[width=7cm]{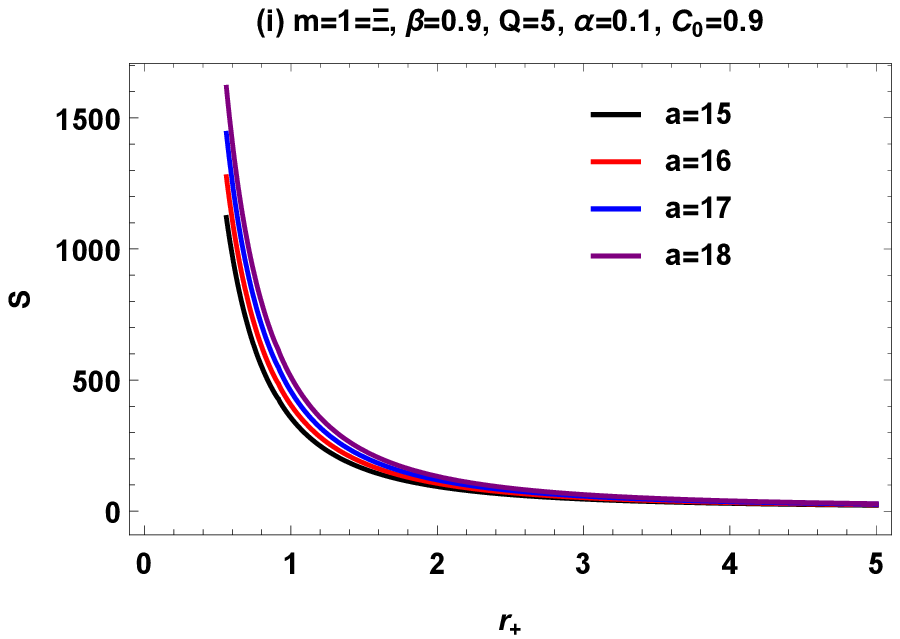}\includegraphics[width=7cm]{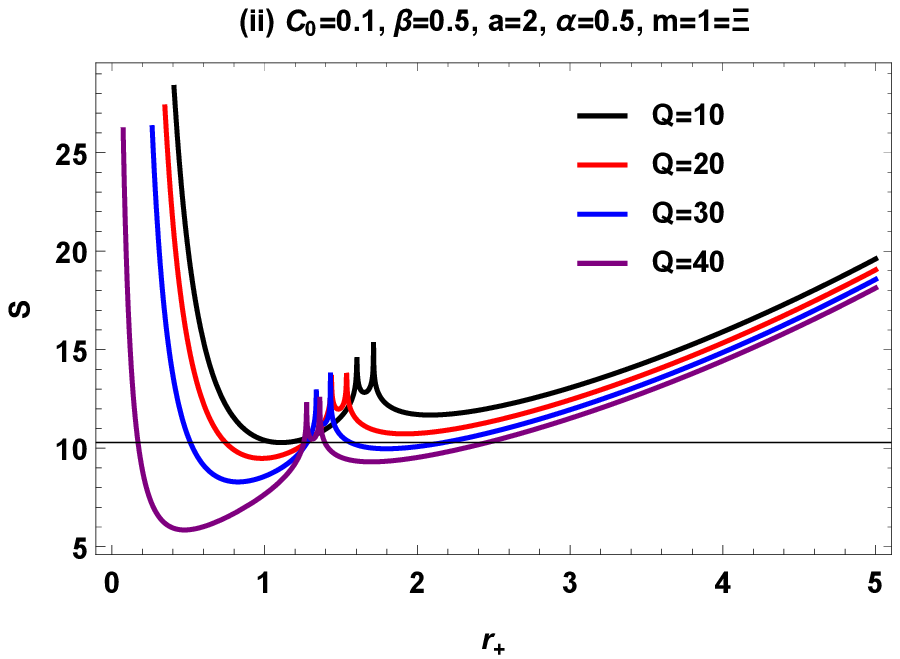}\\
{Figure 3: Entropy versus Horizon for fixed $m=1=\Xi$.}
\end{center}
\textbf{Figure 3}: (i) gives the conduct of $S$ for constant $\beta=0.9=C_{0}$, $\alpha= 0.1$, $Q=5$ and different variations of gravity parameter $\alpha$ in the region $0\leq r_{+}\leq 5$. The $S$ exponentially decreases with the increasing horizon upto $r_+\rightarrow\infty$. This kind of conduct states the stability of BH and the entropy increases within increase in $a$.\\
(ii) depicts the interpretation of entropy versus $ r_{+}$ for
different variations of charge $Q$ and fixed $\beta=0.5=\alpha$,
$C_{0}=0.1$,~$a=2$. At first, the entropy slowly decreases and after
getting a minima, it again starts to increase upto
$r_+\rightarrow\infty$. The entropy decreases for increasing values
of charge $Q$.

\begin{center}
\includegraphics[width=7cm]{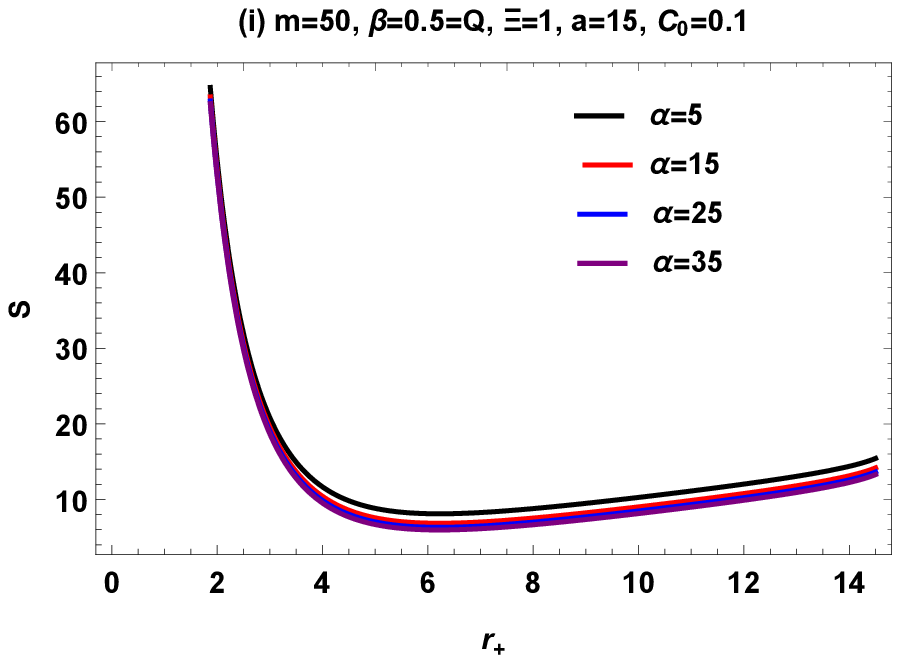}\includegraphics[width=7cm]{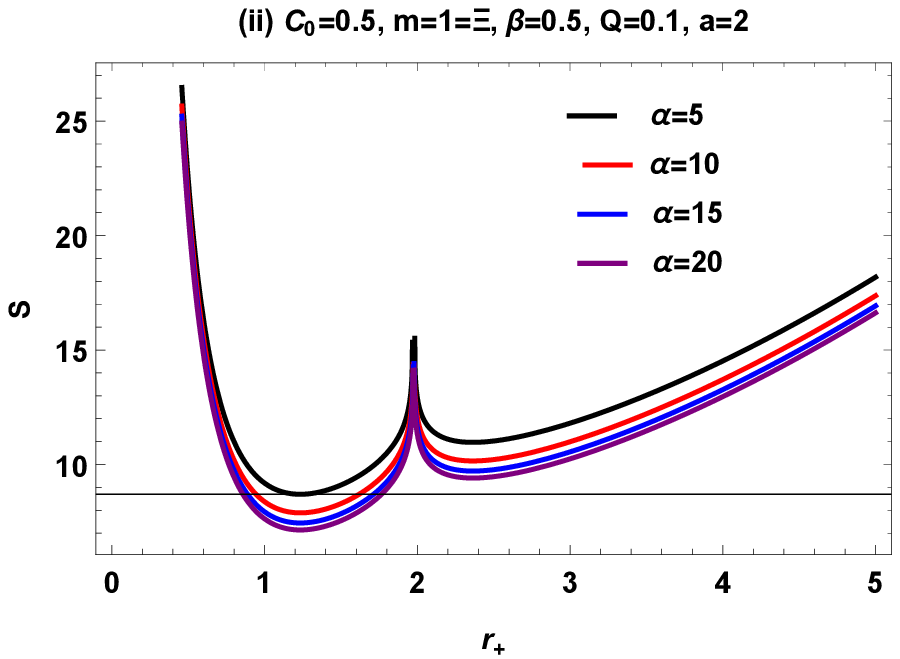}\\
{Figure 4: Entropy w.r.t Horizon for fixed $\Xi=1$ and $\beta=0.5$.}
\end{center}
\textbf{Figure 4}: (i) represents the graphical conduct of $S$ versus horizon for $Q=0.5$, $m=50$, $a=15$, $C_{0}=0.1$ and for various values of $\alpha$ in the range $0\leq r_{+}\leq 15$. It is observable that the entropy slowly decreases for the increasing horizon and after some time it again starts to increase till $r_+\rightarrow\infty$. This condition with positive entropy indicates the stability of BH. The entropy decreases for increasing values of correction parameter $\alpha$.\\
(ii) indicates the role of entropy for $m=1$, $Q=0.1$, $a=2$, $C_{0}=0.5$ and for different variations of $\alpha$ in the domain $0\leq r_{+}\leq 5$. Initially, the entropy decreases and then after attaining a minima it again starts to increase till $r_+\rightarrow\infty$. The initial behavior of entropy looks like a half cardioid.\\
It can be observe from both plots, the entropy is more stable in the
large region of horizon $0\leq r_{+}\leq 15$ as compared to small
radii $0\leq r_{+}\leq 5$ of BH for greater mass $m=50$.

\section{Summary and Discussion}
This paper analyzed the RENLMY BH solution by applying the
Janis-Newman algorithmic rule and complex calculations. Through the
complex computation, we explored the Yukawa BH. If $a\rightarrow 0$,
we obtained the solution of Yukawa BH without Newman-Janis
algorithm. We also investigated the physical properties of BH (i.e.,
$T_{H}$) and analyzed the physical state of RENLMY BH via graphical
interpretation of $T_{H}$ with horizon. The $T_{H}$ depends on the
BH charge $Q$, particle mass $m$ and spin parameter $a$. We have
checked the spin parameter effects on $T_{H}$. The $T_{H}$ decreases
with the increasing spin parameter values. Moreover, we have
investigated the $T'_{H}$ for RENLMY BH with the help of vector
particles tunneling strategy by utilizing the Hamilton Jacobi
strategy. For this investigation, we have utilized the wave equation
of motion with the setting of quantum gravity parameter to study the
vector tunneling of boson particles from RENLMY BH. In the wave
equation, we have applied the WKB approximation that gives a set of
field equations, and after this, by using the separation of
variables technique, we have computed the field equations. The
imagination can be found by using the coefficients of matrix, whose
matrix determinant is equal to zero. We have formulated the
tunneling probability and $T'_{H}$ for the corresponding BH at the
outer horizon by applying surface gravity. The $T'_{H}$ of RENLMY BH
depends upon gravity parameter $\alpha$, BH charge $Q$, particle
mass $m$, spin parameter $a$. It has worth to point out that the
both back-reaction and self-gravitating effects of particles on this
Yukawa BH have been ignored, the computed $T'_{H}$ are the
parameters of Yukawa BH in Newman-Janis algorithm and quantum
gravity. Furthermore, we have discussed the graphical interpretation
of $T'_{H}$ for RENLMY BH. We have analyzed the quantum gravity
effects and spin parameter on $T'_{H}$. We have concluded that the
$T'_{H}$ increases for the increasing values of both correction and
rotation parameter. The $T'_{H}$ at largest value with non-zero
horizon depicts BH leftover mass. After largest values of the
$T'_{H}$ eventually drops down and obtain an asymptotically flat
state till $r_+\rightarrow\infty$, that checks the stable BH
condition. The graphical interpretation of $T'_{H}$ with/without
gravity parameter satisfy the Hawking's strategy (the BH radius size
reduces with the emission of high radiations). This behavior can be
observed in both (Fig. \textbf{1} and \textbf{2}). If $\alpha=0$ in
Eq. (\ref{th2}), we obtain the $T_{H}$ of the Eq. (\ref{th1}).

Furthermore, we have computed the logarithmic entropy corrections by
considering the corrected temperature $T'_{H}$ and basic entropy
$\mathbb{S}_{o}$ for RENLMY BH and also checked the effects of
charge, rotation and gravity parameters on entropy. The entropy
increases with the increasing values of rotation parameter and it
decreases with the increasing values of charge and gravity
parameter. Moreover, we have concluded that the entropy is more
stable in the large region of horizon $0\leq r_{+}\leq 15$ as
compared to small radii $0\leq r_{+}\leq 5$ of BH for greater mass
$m=50$.

\end{document}